\documentclass[letterpaper, 10pt, conference]{IEEEtran}
\IEEEoverridecommandlockouts
\usepackage[centertags]{amsmath}
\usepackage{amsfonts,amssymb}
\usepackage{amsthm}
\usepackage{float}
\usepackage{enumerate}
\usepackage{graphicx}
\usepackage{tikz}
\usepackage{import}
\usepackage[hidelinks]{hyperref}
\usepackage{cleveref}
\usepackage[colorinlistoftodos]{todonotes}
\usepackage{easyReview}
\allowdisplaybreaks
\usetikzlibrary{shapes,arrows,decorations.markings}

\title{\LARGE \bf
Lyapunov-like functions for attitude control via feedback integrators
}

\author{Tejaswi K. C.$^{3}$, Srikant Sukumar$^{1}$ and Ravi Banavar$^{2}$%
\thanks{$^{1}$Associate Professor, Systems and Control Engineering, Indian Institute of Technology, Bombay
        {\tt\small srikant@sc.iitb.ac.in}}%
\thanks{$^{2}$Professor, Systems and Control Engineering, Indian Institute of Technology, Bombay
        {\tt\small ravi.banavar@iitb.ac.in}}%
\thanks{$^{3}$Undergraduate Student, Aerospace Engineering, Indian Institute of Technology, Bombay
        {\tt\small kctejaswi999@gmail.com}}%
}

\newtheorem{theorem}{Theorem}

\newtheorem{remark}{Remark}
\newtheorem{definition}{Definition}

\begin{document}

\maketitle
\thispagestyle{empty}
\pagestyle{empty}

\begin{abstract}
The notion of feedback integrators permits Euclidean integration schemes for dynamical systems evolving on manifolds. Here, a constructive Lyapunov function for the attitude dynamics embedded in an ambient Euclidean space has been proposed. We then combine the notion of feedback integrators with the proposed Lyapunov function  to obtain a feedback law for the attitude control system. The combination of the two techniques yields a domain of attraction for the closed loop dynamics, where earlier contributions were based on linearization ideas. Further, the analysis and synthesis of the feedback scheme is carried out entirely in Euclidean space. The proposed scheme is also shown to be robust to numerical errors.

\end{abstract}

\section{Introduction}

There are many established techniques for attitude control design employing parametrization of the set of rotational matrices~\cite{tsiotras1995new}. A brief summary of the representations involved in description of the kinematics of motion is given in~\cite{shuster1993survey}. Simple control laws in terms of Euler parameters~\cite{mortensen1968globally}, Cayley-Rodrigues parameters~\cite{junkins1991near} have been formulated. However, using such parametrization and hence local charts could cause undesirable unwinding behavior~\cite{bhat2000topological} and require switching between these local coordinate systems for control design.

On the other hand, in the recent past, coordinate-free techniques using geometric ideas have been used to design rigid body attitude controllers~\cite{bloch2003nonholonomic,bullo2004geometric,bayadi2014almost}. However, implementing such feedback laws from geometric control theory~\cite{crouch1984spacecraft,lee2011geometric} requires special variants of numerical integrators (e.g. variational integrator) to preserve the geometric structure of the manifold and yield reliable results.

%In the work of \cite{chang2018controller}, instead of using multiple local charts on the manifold $ \mathcal{M} $ or coordinate-free tools from differential geometry, the system is embedded in an ambient Euclidean space $ \mathbb{R}^n $ and a single coordinate system is used. Then nonlinear controllers can be designed in Euclidean space utilizing existing techniques~\cite{khalil,non_adaptive}.

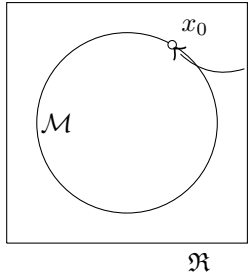
\begin{figure}[h]
	\centering
\begin{tikzpicture}[scale=0.8]
\draw (0, 0) -| (4, 4) node[pos=0.4, below]{$ \mathfrak{R} $} -| (0, 0);
\begin{scope}[shift={(2, 2)}]
\draw (0, 0) circle (1.5cm);
\node[left] at (-1.5/2, 0) {$ \mathcal{M} $};
\path (1.5, 0) arc (0:60:1.5cm) node (eqm) {};
\filldraw[fill=white] (eqm) circle (2pt) node[anchor=south west] {$ x_0 $};
\draw[decoration={markings,mark=at position 0 with {\arrow[scale=2]{<}}},postaction={decorate}]
	(eqm) to[bend right] +(1.2, -0.4);
\end{scope}
\end{tikzpicture}
	\caption{Stabilization in Euclidean space}
	\label{fig:patching}
\end{figure}

While simple geometric PD controllers can be used to stabilize a rigid body~\cite{bullo2004geometric}, numerical integration errors quickly creep into the digital implementations of these schemes, thus resulting in the states not lying on the $ SO(3)$ manifold, and being pushed into the ambient space of $ 3 \times 3 $ real-matrices. In such situations can we still guarantee that these numerical schemes will recover and converge to the manifold? Feedback integrators~\cite{chang_feedback} provide a positive answer to this question. \Cref{fig:patching} illustrates this scenario with $ \mathfrak{R} $ being the set of tuples of $ 3 \times 3 $ matrices and angular velocity vectors, while $ \mathcal{M} $ is the set of tuples of valid rotation matrices and angular velocities. In \cite{chang_feedback} the authors have shown that if the rigid body dynamics is seen as the restriction of a special vector field in an ambient Euclidean space, then Euclidean numerical integration schemes also lead to convergence of states to the manifold. Further, for the case when trajectories starting from an ambient space converge to an embedded  submanifold, \cite{arsie_patching} shows that the omega limit set lies in a unique connected component of the level sets corresponding to a Lyapunov-like function. Our work builds on these two techniques to design Euclidean controllers which guarantee that the rigid body converges to an equilibrium point $ x_0 $ on $ \mathcal{M} $, even if at some instants the states do not lie on $ \mathcal{M} $.

More recent work by \cite{chang_controller} has addressed this problem by linearizing the ambient dynamics, 
thus is only valid in a small neighborhood around the desired set-point. We briefly introduce the same in \cref{sec:chang}. In this article (primarily in \cref{sec:nonlinear}), we have developed a new procedure for nonlinear design using Lyapunov-like functions on the ambient system to guarantee asymptotic convergence to an equilibrium point in $ \mathcal{M} $. Finally, to demonstrate the performance of the controller, numerical simulations are presented in \cref{sec:simulations}.

\subsection*{Notation:}
\begin{itemize}
\item Euclidean inner product is used in this paper:
\[ \langle A, B\rangle = \sum_{i,j} A_{i,j}B_{i,j} = tr(A^T B) \]
for matrices of identical dimensions. The norm induced by this inner product is used for vectors and matrices.
\item $ SO(3) = \lbrace R \in \mathbb{R}^{3\times3} \mid R^T R = I, \det(R) = 1 \rbrace $ is the Lie group of all rotations and $ \mathfrak{so}(3) = \lbrace A \in \mathbb{R}^{3\times3} \mid A = -A^T \rbrace $ is the corresponding Lie algebra.
\item Hat map $ \wedge : \mathbb{R}^3 \to \mathfrak{s0}(3) $,
\[ \hat{\Omega} = \begin{bmatrix}
0 & -\Omega_3 & \Omega_2 \\
\Omega_3 & 0 & -\Omega_1 \\
-\Omega_2 & \Omega_1 & 0
\end{bmatrix} \]
for $ \Omega \in \mathbb{R}^3 $. The inverse map is the vee map, $ \vee $, such that $ (\hat{\Omega})^\vee = \Omega $ for all $ \Omega \in \mathbb{R}^3 $ and $ \widehat{(A^\vee)} = A $ for all $ A \in \mathfrak{so}(3) $.
\item For a square matrix $ A $, $ A_s := (A + A^T)/2 $ is the symmetric part and $ A_k := (A - A^T)/2 $ is the skew-symmetric part.
\end{itemize}

\section{Stabilization of a rigid body using linearization}\label{sec:chang}
This section summarizes the linearization procedure introduced in \cite{chang_controller}. Consider a control system $ \Sigma $ on $ \mathbb{R}^n $,
\begin{equation*}
\Sigma : \dot{x} = X(x, u), x \in \mathbb{R}^n, u \in \mathbb{R}^k
\end{equation*}
Assume that there is an m-dimensional submanifold $ \mathcal{M} $ of $ \mathbb{R}^n $ that is invariant under the flow of the system. So we can restrict the system to $ \mathcal{M} $ as,
\begin{equation*}
\Sigma|\mathcal{M} : \dot{x} = X(x, u), x \in \mathcal{M}, u \in \mathbb{R}^k
\end{equation*}

It is convenient to use the ambient control system $ \Sigma $ and the Cartesian coordinates on the ambient space in order to design controllers for the system $ \Sigma|\mathcal{M} $ on the manifold $ \mathcal{M} $.

Let $ \tilde{V} $ be a non-negative function on the euclidean space such that $ \mathcal{M} = \tilde{V}^{-1}(0) $. At every point in $ \mathcal{M} $ as $ \tilde{V} $ attains its minimum value of $ 0 $, $ \nabla\tilde{V}(x) = 0, \forall x \in \mathcal{M} $. We obtain a new ambient control system by subtracting $ \nabla\tilde{V} $ from the control vector field,
\begin{equation*}
\tilde{\Sigma} : \dot{x} = \tilde{X}(x, u), x \in \mathbb{R}^n, u \in \mathbb{R}^k
\end{equation*}
with $ \tilde{X}(x, u) = X(x, u) - \nabla\tilde{V}(x) $. It is easily verified that $ \tilde{\Sigma}|\mathcal{M} = \Sigma|\mathcal{M} $, meaning that the system dynamics is preserved on $ \mathcal{M} $. The negative gradient of $ \tilde{V} $ helps in making $ \mathcal{M} $ attractive for $ \tilde{\Sigma} $ dynamics \cite{chang_feedback}.

Now, let $ (x_0, u_0) \in \mathcal{M} \times \mathbb{R}^k $ be an equilibrium point of $ \Sigma|\mathcal{M} $ with $ X(x_0, u_0) = 0 $. Jacobian linearization can be carried out on the ambient system $ \tilde{\Sigma} $ around the equilibrium point in the ambient space to come up with stabilizing controllers for the original system on the manifold. The linearization of $ \tilde{\Sigma} $ is given by,
\begin{equation*}
\tilde{\Sigma}_0^l : \dot{x} = \frac{\partial \tilde{X}}{\partial x}(x_0, u_0)(x - x_0) + \frac{\partial \tilde{X}}{\partial u}(x_0, u_0)(u - u_0)
\end{equation*}
where $ (x, u) \in \mathbb{R}^n \times \mathbb{R}^k $.

\begin{theorem}\cite[Theorem II.3]{chang_controller}\label{thm:linear_stab}
If a linear feedback controller $ u : \mathbb{R}^n \to \mathbb{R}^k $ exponentially stabilizes the equilibrium point $ x_0 $ for the linearization $ \tilde{\Sigma}_0^l $ of the ambient system $ \tilde{\Sigma} $, then it also exponentially stabilizes the equilibrium point $ x_0 $ for $ \Sigma|\mathcal{M} $.
\end{theorem}

We are concerned the application of \cref{thm:linear_stab} to the \textit{rigid body system} with full actuation,
\begin{equation} \label{eq:rigid_body}
\begin{aligned}
\dot{R} &= R \hat{\Omega} \\
\dot{\Omega} &= u
\end{aligned}
\end{equation}
where $ (R, \Omega) \in \mathcal{M} \subset \mathbb{R}^{3 \times 3}\times \mathbb{R}^3 $. $ \mathcal{M} = SO(3) \times \mathbb{R}^3$ is the invariant manifold being considered. It is assumed that the control input is appropriately scaled and shifted to account for nonlinear terms in the dynamics.

Let $ GL^{+}(3) = \lbrace R \in \mathbb{R}^{3 \times 3} \mid \det{R} > 0 \rbrace $ and define a function $ \tilde{V} $ on $ GL^{+}(3) \times \mathbb{R}^3 $ by
\begin{equation}\label{eq:V_tilde}
\tilde{V}(R, \Omega) = \frac{k_e}{4} ||R^T R - I ||^2
\end{equation}
with constant $ k_e > 0 $. One can verify that $ \tilde{V}^{-1}(0) = \mathcal{M} $ and
\[ \nabla_R\tilde{V} = -k_e R(R^TR-I), \nabla_\Omega \tilde{V} = 0 \]
So the modified rigid body system ($ \tilde{\Sigma} $) in the ambient space is,
\begin{equation} \label{eq:rigid_body_attr}
\begin{aligned}
\dot{R} &= R \hat{\Omega} - k_e R (R^T R - I) \\
\dot{\Omega} &= u
\end{aligned}
\end{equation}

To design a controller, the system~\eqref{eq:rigid_body_attr} is linearized to get,
\begin{equation}\label{eq:rigid_body_linear}
\begin{aligned}
\dot{Z}_s &= -2 k_e Z_s \\
\dot{Z}_k^{\vee} &= \Omega \\
\dot{\Omega} &= u
\end{aligned}
\end{equation}
with $ Z = R_0^T \Delta R = R_0^T (R - R_0) $ being a transformation of $ R $.

For $ \mathbb{R} \ni k_p, k_d > 0 $, the linear PD controller
\begin{equation*}
u = -k_p Z_k^\vee - k_d \Omega
\end{equation*}
exponentially stabilizes the equilibrium point $ (R_0, 0) $ for both the linearized system \eqref{eq:rigid_body_linear} and the rigid body system \eqref{eq:rigid_body} on $ \mathcal{M} $.

\section{Ambient control formulation using Lyapunov-like functions}\label{sec:nonlinear}
The results of the previous section are obtained via linearization and therefore suffer from obvious drawbacks such as the inability to accurately estimate the region of convergence. In this section, we present a novel nonlinear design method based on Lyapunov techniques which is utilized to stabilize the rigid body using feedback integrators.

An important result on locating $ \omega $-limit sets using height functions is employed. Given the bounded solution of an autonomous vector field on a Riemannian manifold, a finer estimate of the location of $ \omega $-limit set can be obtained using results in \cite{arsie_patching}. We summarize the same here.

\subsection{Preliminaries}\label{sec:prelim}
The set-up in \cite[Section 2]{arsie_patching} is restated while changing the notations from $ \mathcal{M} $ to $ \mathfrak{R} $, $ S $ to $ \mathcal{M} $ and $ \Omega $ to $ \omega $ so that consistency with the rest of our paper is maintained:
\begin{itemize}
\item A Riemannian manifold $ (\mathfrak{R}, g) $ of class $ C^2 $ on which a locally Lipschitz continuous vector field
\begin{equation}\label{eq:vec_field}
\dot{x} = f(x)
\end{equation}
is given.
\item Consider a Cauchy problem for \eqref{eq:vec_field} with initial value $ x(0) $ such that the corresponding solution $ x(t, x(0)) $ is bounded.
\item Assume that the $ \omega $-limit set $ \omega(x(0)) $, which is a compact and connected set, is contained in a closed embedded submanifold $ \mathcal{M} \subset \mathfrak{R} $. Equivalently $ \mathcal{M} $ is attracting for the solution of \eqref{eq:vec_field} starting at $ x(0) $.
\item Let $ O $ be an open tubular neighborhood of $ \mathcal{M} $ in $ \mathfrak{R} $. Assume that there exists a real-valued $ C^1 $ function $ W : O \to \mathbb{R} $ such that $ \dot{W}(x) \le 0 $ on $ \mathcal{M} $, where $ \dot{W}(x) $ is the derivative of $ W(x) $ along the flow (Lie derivative). Moreover, let $ E := \lbrace x \in \mathcal{M} \mid \dot{W}(x) = 0 \rbrace $ so that $ \dot{W}(x) < 0 $ on $ \mathcal{M} \setminus E $.
\end{itemize}

The function $ W $ as described above is called a \textit{height function} for the pair $ (\mathcal{M}, f) $.

\begin{definition}\cite[Definition 5]{arsie_patching}\label{def:containment}
Let $ \lbrace E_i \rbrace_{i \in \mathbb{I}} $ be the connected components of $ E $, where $ \mathbb{I} = \lbrace 1, 2, \dots \rbrace \subset Z^+ $ . Given a function $ W $ as in the assumptions, we say that the components $ \lbrace E_i \rbrace_{i \in \mathbb{I}} $ are \emph{contained} in $ W $ if each $ E_i $ lies in a level set of $ W $, and the subset $ \lbrace W(E_i) \rbrace_{i \in \mathbb{I}} \subset \mathbb{R} $ has at most a finite number of accumulation points in $ \mathbb{R} $.
\end{definition}

The main result is stated below.
\begin{theorem}\cite[Theorem 6]{arsie_patching}\label{thm:containment}
If the components $ \lbrace E_i \rbrace_{i \in \mathbb{I}} $ are \emph{contained} in $ W $ according to \cref{def:containment}, then $ \omega(x(0)) \subset E_i $ for a unique $ i \in \mathbb{I} $.
\end{theorem}

\subsection{Rigid body stabilization}
Using the above result, we propose an ambient nonlinear controller for rigid body stabilization. Consider again the feedback integrator form of the rigid body dynamics \eqref{eq:rigid_body_attr},
\begin{equation}\label{eq:rigid_body_attr_copy}
\begin{aligned}
\dot{R} &= R \hat{\Omega} - k_e R (R^T R - I) \\
\dot{\Omega} &= u
\end{aligned}
\end{equation}
For this system, $ \mathcal{M} = SO(3) \times \mathbb{R}^3$ is the invariant manifold being considered which is embedded in the ambient space $ \mathfrak{R} = \mathbb{R}^{3 \times 3} \times \mathbb{R}^3 $. Also, consider again the function $ \tilde{V} $ defined in \eqref{eq:V_tilde}.

\begin{theorem}\label{thm:rigd_body_stab}
The control law given by,
\begin{equation}\label{eq:controller}
u = -k_p Z_k^\vee - k_d \Omega, \quad \mathbb{R} \ni k_p, k_d > 0
\end{equation}
asymptotically stabilizes an equilibrium point $ (R_0, 0) \in \mathcal{M} $ of the system \eqref{eq:rigid_body_attr_copy} for almost all initial conditions starting from $ \tilde{V}^{-1}([0, c]) $ and some $ c > 0 $.
\end{theorem}

\noindent The corresponding closed-loop system employing \eqref{eq:rigid_body_attr_copy} and \eqref{eq:controller} is,
\begin{equation} \label{eq:rigid_body_euc}
\begin{aligned}
\dot{R} &= R \hat{\Omega} - k_e R (R^T R - I) \\
\dot{\Omega} &= -k_p Z_k^\vee - k_d \Omega
\end{aligned}
\end{equation}
with $ (R, \Omega) $ as its states.
\begin{proof}
We first verify the assumptions corresponding to the set-up in \cref{sec:prelim},
\begin{itemize}
\item We have a Riemannian manifold $ (\mathbb{R}^{3\times 3} \times \mathbb{R}^3, \cdot ) $ on which a locally Lipschitz continuous vector field \eqref{eq:rigid_body_euc} is given.
\item It can be directly claimed from \cite[Theorem 2]{chang2018controller} that every trajectory of \eqref{eq:rigid_body_attr_copy} starting from a point in $ \tilde{V}^{-1}([0, c]) $, for some $ c > 0 $, stays in $ \tilde{V}^{-1}([0, c]) $ for all $ t \ge 0 $ and asymptotically converges to the set $ \mathcal{M} =  \tilde{V}^{-1}(0) $ as $ t \to \infty $. Since $ \tilde{V}^{-1}([0, c]) $ is compact and positively invariant, the first state $ R $ in \eqref{eq:rigid_body_euc} is bounded if initial states $ (R(0), \Omega(0)) \in \tilde{V}^{-1}([0, c]) $. Now, consider a function $ V_2 = \frac{1}{2} ||\Omega||^2 $ whose derivative is evaluated using \eqref{eq:rigid_body_euc}:
{\small \begin{gather*}
\dot{V}_2 = - k_d ||\Omega||^2 - k_p \Omega^T Z_k^\vee \\
\le - k_d ||\Omega|| \left( ||\Omega|| - \frac{k_p}{k_d} ||Z_k^\vee|| \right)
\le - k_d \epsilon ||\Omega||^2 \le 0
\end{gather*}}
if $ \displaystyle ||\Omega|| \ge \frac{k_p}{k_d (1-\epsilon)} ||Z_k^\vee|| $. We know that $ Z_k^\vee $ is bounded because $ R $ is already shown to be bounded. Hence, either $ ||\Omega|| $ is bounded by a fraction of $ ||Z_k^\vee|| $ or $ \dot{V}_2 \le 0 $ implying that $ ||\Omega|| $ is non-increasing. So, $ \Omega $ is bounded.

Therefore we have a Cauchy problem for \eqref{eq:rigid_body_euc} with initial value $ (R(0), \Omega(0)) $ such that the solution is bounded.
\item From the previous point, we know that the $ \omega $-limit set $ \omega(R(0), \Omega(0)) $, which is a compact and connected set, is contained in a closed embedded submanifold $ \mathcal{M} = SO(3) \times \mathbb{R}^3  \subset \mathfrak{R} = \mathbb{R}^{3\times 3} \times \mathbb{R}^3 $.
\item Let $ O $ be an open tubular neighborhood of $ \mathcal{M} $ in $ \mathfrak{R} $. This set is being used in our context to help determine the region of convergence in $ \mathfrak{R} $. There is a real-valued $ C^1 $ function $ W : O \to \mathbb{R} $ such that $ \dot{W} \le 0 $ on $ \mathcal{M} $, defined as below:
\begin{equation}\label{eq:height_func}
W(R, \Omega) = \frac{k_p}{4} (||Z_s||^2 + ||Z_k||^2) + \frac{1}{2}||\Omega||^2 + \epsilon \langle Z_k^\vee, \Omega \rangle
\end{equation}
which serves as the height function with $ Z = R_0^T (R - R_0) $. The derivative of $ W(R, \Omega) $ along the flow on $ \mathcal{M} $ is (\cref{sec:deriv_height}),
{\small \begin{equation}\label{eq:height_deriv}
\dot{W}|_\mathcal{M}(R, \Omega) \le -(k_d-\epsilon) ||\Omega||^2 - \epsilon k_d \langle Z_k^\vee, \Omega \rangle - \epsilon k_p ||Z_k^\vee||^2 \le 0
\end{equation}}
for $\displaystyle 0 < \epsilon < \frac{4k_pk_d}{4k_p+k_d^2} $.

$ E $ is defined as $ \lbrace (R, \Omega) \in \mathcal{M} \mid \dot{W}(R, \Omega) = 0 \rbrace $:
\begin{equation}\label{eq:E_zero}
E = \lbrace (R, \Omega) \in SO(3) \times \mathbb{R}^3 \mid Z_k^\vee = 0, \Omega = 0 \rbrace
\end{equation}
so that $ \dot{W}(R, \Omega) < 0 $ on $ \mathcal{M} \setminus E $.
\end{itemize}

\noindent Thus the main assumptions required for \cref{thm:containment} are satisfied.

We observe that the height function can be re-written as
{\small \[ W(R, \Omega) = \frac{k_p}{4} tr((I - R_0^T R)^T(I - R_0^T R)) + \frac{1}{2}||\Omega||^2 + \epsilon \langle Z_k^\vee, \Omega \rangle \]}
which is different from standard Lyapunov functions used for rigid body stabilization like $ V(R, \Omega) = \frac{k_p}{4} tr(I - R_0^T R) + \frac{1}{2}||\Omega||^2 $. Among other changes, it has an additional cross term $ \langle Z_k^\vee, \Omega \rangle $ which helps us to identify the equilibrium point as one of the connected components of $ E $ and then employ \cref{thm:containment} to prove asymptotic convergence.

\begin{remark}\label{rem:neighborhood}
The maximum value that the real number $ c $ can take is less than $ k_e/ 12 $ (see \cref{sec:c_value}) and this ensures that there exists an open tubular neighborhood $ O $ which is the superset of $ \tilde{V}^{-1}([0, c)) $.
\end{remark}

\noindent On the set $ E $ we know that,
\begin{gather*}
Z_k^\vee = 0 \Rightarrow R_0^T R - R^T R_0 = 0
\end{gather*}
With $ \tilde{R} = R_0^T R $,
\[ \tilde{R} = \tilde{R}^T \Rightarrow \tilde{R}^2 = I \]
Using the axis-angle representation of rotation matrices~\cite{murray2017mathematical}, $ \tilde{R} = \exp(\theta \hat{\xi}) = I + sin{\theta}\hat{\xi} + (1 - cos{\theta})\hat{\xi}^2, \hat{\xi} \in \mathfrak{so}(3) $,
\begin{gather*}
\tilde{R}^2 = I \Rightarrow e^{2\theta\hat{\xi}} = I = e^{2n\pi\hat{k}} \quad k, \xi \in \mathbb{R}^3, ||k|| = 1 \\
\Rightarrow \theta = n\pi, \hat{\xi} = \hat{k}
\end{gather*}
The set of all such matrices, $ \tilde{R} $, can be divided into two sets as follows,
\begin{gather}
\theta = 2 m \pi, m \in \mathbb{Z} \Rightarrow \tilde{R} = I \Rightarrow tr(\tilde{R}) = 3 \\
\begin{aligned}\label{eq:Zkvee_zero}
\theta &= (2 m + 1) \pi, m \in \mathbb{Z} \\
\Rightarrow \tilde{R} &= \exp(\pi\hat{\xi}) \neq I \Rightarrow tr(\tilde{R}) = 1 + 2\cos{\theta} = -1
\end{aligned}
\end{gather}
Thus $ E = E_1 \cup E_2 $ is described below,
\begin{itemize}
\item  $ E_1 = \lbrace (R_0, 0) \rbrace $. As this subset contains only one point, it is trivially connected. Value of $ W $ in $ E_1 $ evaluated with $ (Z, \Omega) = (0,0) $ gives $ W(R, \Omega) = 0 $.
\item $ E_2 = \lbrace (R, \Omega) \in SO(3) \times \mathbb{R}^3 \mid tr(R_0^T R) = tr(\tilde{R}) = -1, \tilde{R} = \tilde{R}^T, \Omega = 0 \rbrace $. A point in the set $ E_2 $ has the form $ (R_0 \tilde{R}, 0) $ where $ \tilde{R} = e^{\pi \hat{\xi}} $ for an unit vector $ \xi $ as in \eqref{eq:Zkvee_zero}. Consider two points $ x_1 = (R_0 e^{\pi \hat{\xi}_1}, 0)\ \text{and}\ x_2 = (R_0 e^{\pi \hat{\xi}_2}, 0) $ in $ E_2 $ and the corresponding axis vectors $ \xi_1, \xi_2 \in \mathbb{R}^3 $ with unit magnitudes. Define a path variable $\displaystyle \xi(\alpha) = \frac{(1-\alpha) \xi_1 + \alpha \xi_2}{||(1-\alpha) \xi_1 + \alpha \xi_2||}, \alpha \in [0, 1] $ such that $ \xi(0) = \xi_1 $ and $ \xi(1) = \xi_2 $. The corresponding path in $ \mathcal{M} $ connecting $ x_1 $ and $ x_2 $ is $ \lbrace x(\alpha) = (R(\alpha), \Omega(\alpha)) \in SO(3) \times \mathbb{R}^3 \mid R(\alpha) = R_0 \tilde{R}(\alpha) = R_0 e^{\pi \widehat{\xi(\alpha)}}, \Omega=0 \rbrace $. Any point $ x(\alpha) $ in this path connecting $ x_1, x_2 $ also belongs to $ E_2 $, meaning the set $ E_2 $ is path connected and hence connected.
Evaluating $ W $ in $ E_2 $ with $ (Z_k, \Omega) = (0,0) $, 
\begin{multline*}
W(R, \Omega) = \frac{k_p}{4}||Z_s||^2 = \frac{k_p}{4} tr((R_0^TR - I ) (R_0^TR - I ))\\
= \frac{k_p}{4} tr(\tilde{R}^2 - 2\tilde{R} + I) = \frac{k_p}{4} tr(2I - 2\tilde{R}) = 2k_p
\end{multline*}
as $ \tilde{R} = R_0^T R = \tilde{R}^T $ and $ tr(R_0^T R) = -1 $ on $ E_2 $.
\end{itemize}

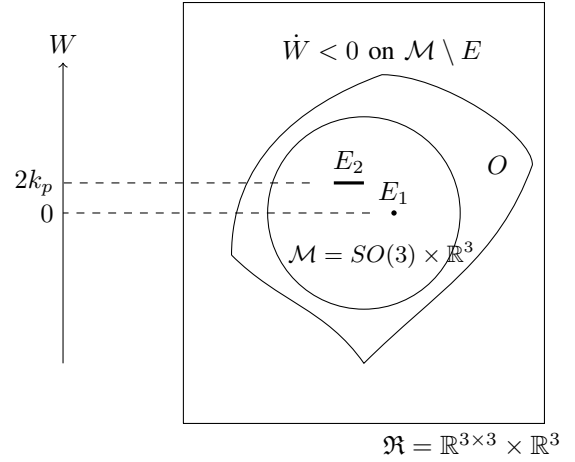
\begin{figure}[h!]
	\centering
\begin{tikzpicture}[scale=0.8]
%\draw [help lines, very thin] (-2, -2) grid (8,5);
\begin{scope}[shift={(-2, 0)}]
\draw [->] (0,-1) -- (0,4) node[above] {$ W $};
\end{scope}
\draw (0, -2) -| (6, 5) node[pos=0.4, below]{$ \mathfrak{R} = \mathbb{R}^{3\times 3} \times \mathbb{R}^3 $} -| (0, -2);
\draw (3, 1.5) circle (1.6cm);
\node[left] at (5, 0.8) {{\small $ \mathcal{M} = SO(3) \times \mathbb{R}^3 $}};
%\draw (0.8, 0.8) .. controls +(0, 1.5) .. ++(2, 3) node[above] {$ \dot{W} < 0\ \text{on}\ \mathcal{M} \setminus E $} .. controls +(right:1cm) and +(up:0.5cm) .. ++(2.5, -1.5) .. controls +(0, -1) .. node[pos=0.5, below=0.5cm] {$ \mathcal{M} = SO(3) \times \mathbb{R}^3 $} (3, -1) .. controls +(-1, 0) .. (0.8, 0.8);
\draw (0.8, 0.8) to[out=90, in=200] ++(2.5, 3) node[above] {$ \dot{W} < 0\ \text{on}\ \mathcal{M} \setminus E $} .. controls +(right:1cm) and +(up:0.5cm) .. ++(2.5, -1.5) node[left=0.2cm] {$ O $} to[out=250, in=45] (3, -1)  to[out=125, in=-45] (0.8, 0.8);
\draw[very thick] (2.5, 2) to node[pos=0.5, above] {$ E_2 $} (3, 2);
\draw [dashed] (2.1, 2) -- +(-2.1-2, 0) node [left] {$ 2 k_p $};
\filldraw (3.5, 1.5) circle (1pt) node[above] {$ E_1 $};
\draw [dashed] (3.1, 1.5) -- +(-5.1, 0) node [left] {$ 0 $};
\end{tikzpicture}
	\caption{Illustration of the components involved for the case of rigid body stabilization. A general version can be found in \cite{arsie_patching}.}
	\label{fig:manifold_height}
\end{figure}

\Cref{fig:manifold_height} illustrates the basic components involved in this proof. $ \mathfrak{R} $ is the set of tuples of $ 3 \times 3 $ matrices and angular velocity vectors, while $ \mathcal{M} $ is the set of tuples of valid rotation matrices and angular velocities. Further, $ O $ is the tubular neighborhood of $ \mathcal{M} $ in which the height function $ W $ is defined. On the y-axis, the value of $ W(R, \Omega) $ for any $ (R, \Omega) \in O $ is shown. We have already proved that $ \dot{W}(R, \Omega) < 0 $ on $ \mathcal{M} \setminus (E_1 \cup E_2) $. From \eqref{eq:height_func} and \eqref{eq:E_zero}, any subset of $ E $ which lies in a level set of $ W $ will have the structure,
\[ W^{-1}(\frac{k_p}{4} ||Z_s||^2 = c) \]
where $ c \ge 0 $. As shown earlier, the connected components of $ E $ lie in level sets of $ W $.

\noindent Finally completing the arguments of the proof,
\begin{itemize}
\item $ E_1 \subset W^{-1}(0), E_2 \subset W^{-1}(2k_p) $ which implies that $ \lbrace W(E_i) \rbrace_{i \in {1, 2}} = \lbrace 0, 2k_p \rbrace $ has no accumulation point in $ \mathbb{R} $. Hence, we can say that $ \lbrace E_i \rbrace_{i \in {1,2}} $ are \emph{contained} in $ W $ using \cref{def:containment}.

\item Employing \cref{thm:containment}, $ \omega(R(0), \Omega(0)) \subset E_i $ for a unique $ i \in \lbrace 1, 2 \rbrace $.

\item From the dynamics \eqref{eq:rigid_body_euc} we know that $ E_2 $ is forward invariant, that is,
\begin{gather*}
Z_k^\vee = 0, \Omega = 0, R^T R = I \\
\Rightarrow \dot{R} = 0, \dot{\Omega} = 0
\end{gather*}
on $ E_2 $. In addition, $ E_2 $ is an unstable set for the system dynamics on $ \mathcal{M} $~\cite{bayadi2014almost}.
\end{itemize}

This proves that $ \omega(R(0), \Omega(0)) \subset E_1 = \lbrace (R_0, 0) \rbrace $ for almost all $ (R(0), \Omega(0)) $ in $ \tilde{V}^{-1}([0, c]) $.
\end{proof}

\section{Simulations}\label{sec:simulations}
In this section we look at a few numerical examples to illustrate the strategies previously presented.

\subsection{Ideal case}\label{sec:sim_normal}

\begin{figure}[h!]
\centering
\includegraphics[width=\linewidth]{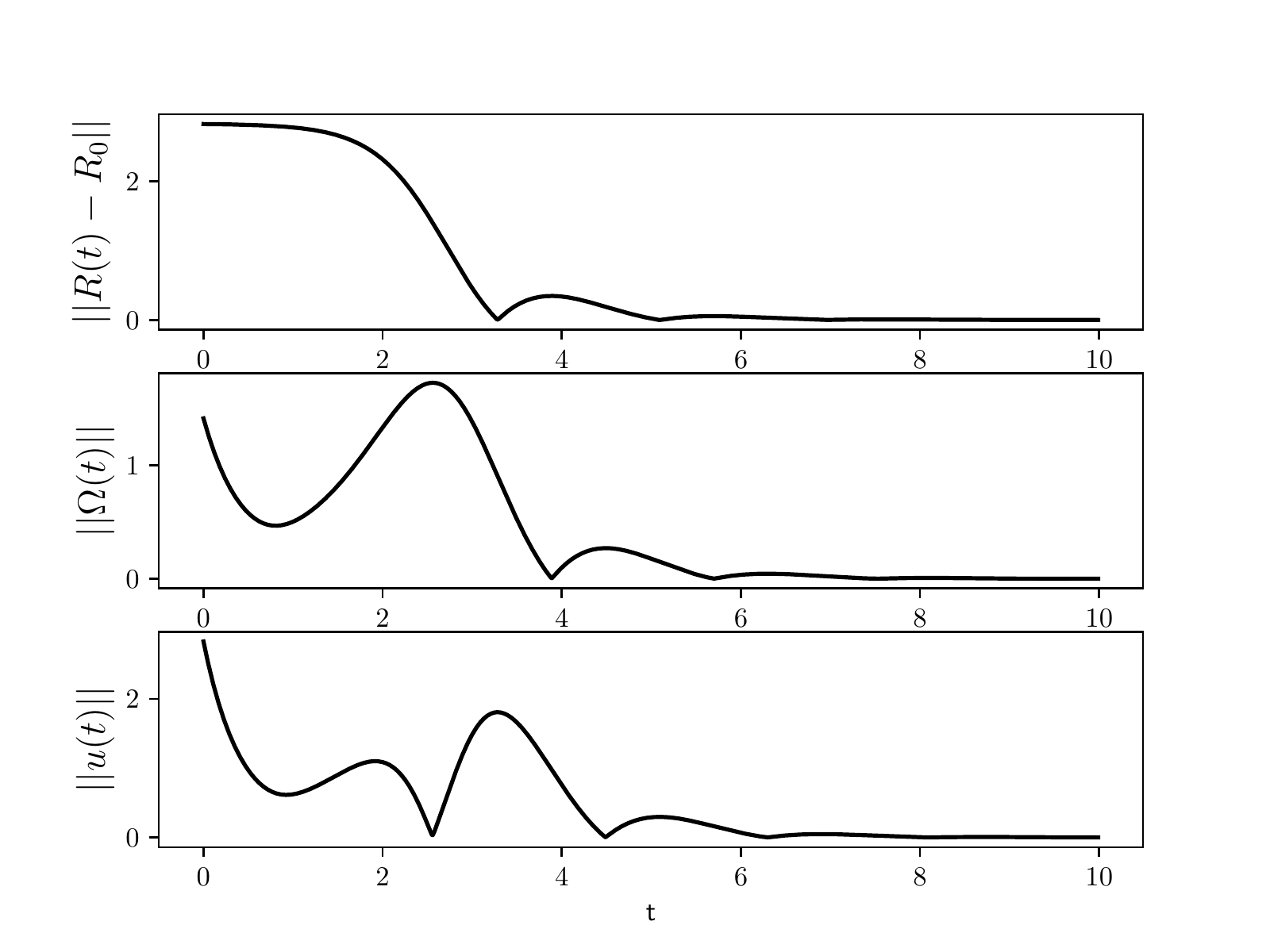}
\caption{Nonlinear rigid body stabilization using height function}
\label{fig:euc_rigid_normal}
\end{figure}

We demonstrate the performance of the rigid body system~\eqref{eq:rigid_body_euc} with initial conditions in the manifold, $ \mathcal{M} $. Consider $ (R_0, 0) $ as the desired equilibrium point ($ R_0 = diag\left\lbrace-1, -1, 1\right\rbrace $) along with the initial conditions,
\begin{gather*}
R(0) = \exp(\frac{2\pi}{3}\hat{e}_2),\quad \Omega(0) = [0, 1, 1]^T, \text{where}\ e_2 = [0,1,0]^T
\end{gather*}
and the parameters being,
\begin{gather*}
k_e = 1, k_p = 4, k_d = 2
\end{gather*}
The chosen value of $ \epsilon $ is $  0.99 \times \frac{4k_pk_d}{4k_p+k_d^2} = 1.584 $ which satisfies the constraint needed in \eqref{eq:height_deriv}.

\Cref{fig:euc_rigid_normal} depicts the magnitudes of orientation and attitude errors along with the control magnitude. We recover the expected ideal performance in this case.

%\subsection{Initial conditions outside the sub-manifold}\label{sec:sim_outside}
%For the purpose of this illustration, consider the initial attitude of the body not on $ SO(3) $:
%\begin{gather*}
%R(0) = \begin{bmatrix}
%-0.55 & 0 & 0.953 \\
%0 & 1.1 & 0 \\
%-0.953 & 0 & -0.55
%\end{bmatrix} = 1.1 \times \exp(\frac{2\pi}{3}\hat{e}_2)
%\end{gather*}
%with all the other conditions and parameters being the same as \cref{sec:sim_normal}. One can verify that,
%\begin{equation*}
%||R(0)^T R(0) - I|| = 0.3637 < \sqrt{1/3}
%\end{equation*}
%meaning the initial condition is in the allowed set (\cref{sec:c_value}).
%
%\begin{figure}[h!]
%\centering
%\includegraphics[width=\linewidth]{images/euc_rigid_outside.pdf}
%\caption{Stabilization with initial conditions outside the sub-manifold}
%\label{fig:euc_rigid_outside}
%\end{figure}
%
%We still observe excellent performance of the system in \cref{fig:euc_rigid_outside} (plotted with a logarithmic scale). The last sub-figure represents the amount of deviation of $ R $ from $ SO(3) $, and it is clearly observed that it goes to zero.

\subsection{Numerical Robustness}\label{sec:sim_noise}
\begin{figure}[h!]
\centering
\includegraphics[width=\linewidth]{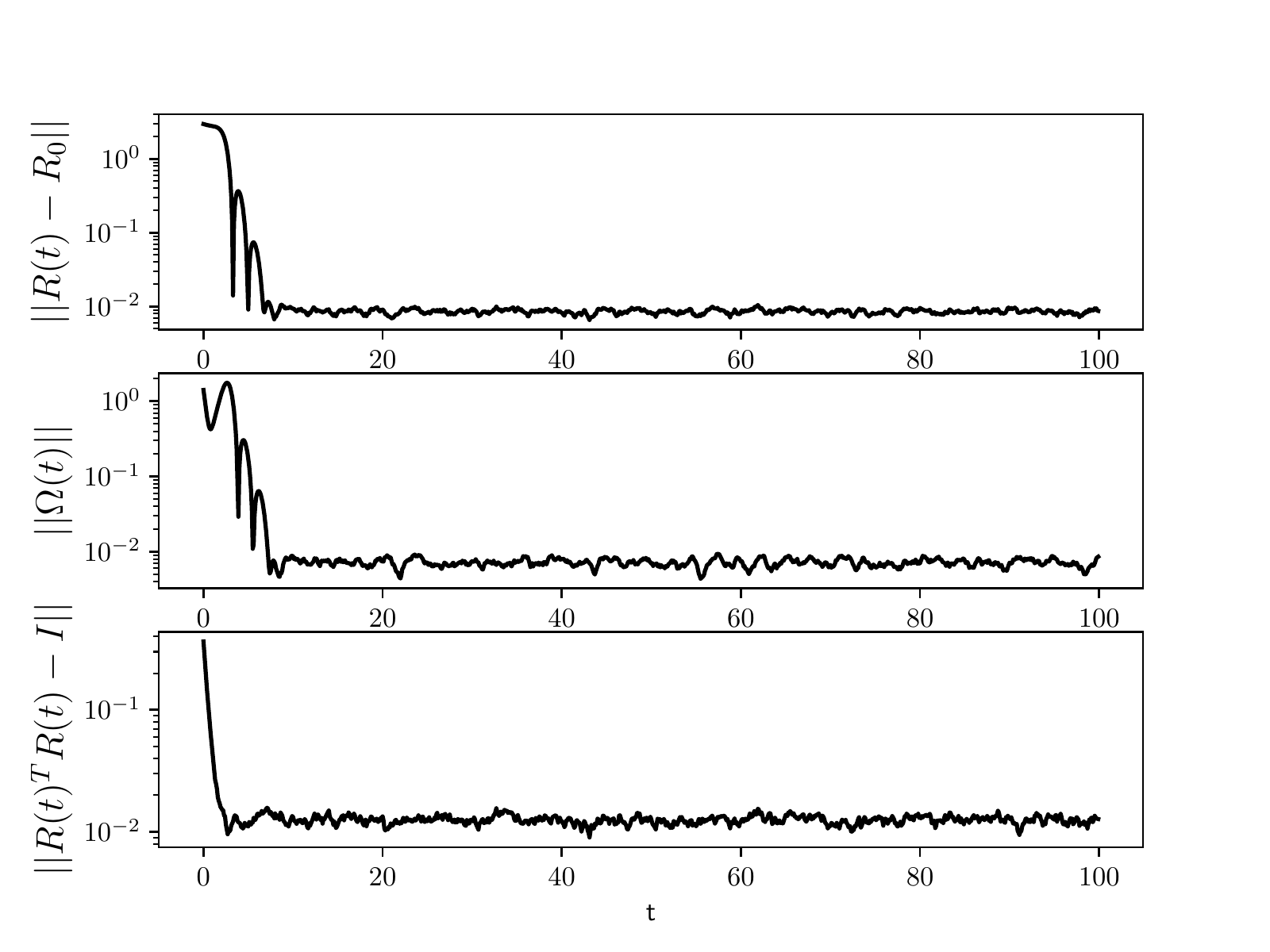}
\caption{Demonstration of robustness}
\label{fig:euc_rigid_noise}
\end{figure}

To illustrate the strength of the proposed control, consider an initial state of the body not on $ \mathcal{M} $:
%\begin{gather*}
%R(0) = \begin{bmatrix}
%-0.55 & 0 & 0.953 \\
%0 & 1.1 & 0 \\
%-0.953 & 0 & -0.55
%\end{bmatrix} = 1.1 \times \exp(\frac{2\pi}{3}\hat{e}_2) \in \mathbb{R}^{3 \times 3} \setminus SO(3)
%\end{gather*}
\begin{gather*}
R(0) = 1.1 \times \exp(\frac{2\pi}{3}\hat{e}_2) \in \mathbb{R}^{3 \times 3} \setminus SO(3)
\end{gather*}
with all the other conditions and parameters being identical to \cref{sec:sim_normal}. One can verify that,
\begin{equation*}
||R(0)^T R(0) - I|| = 0.3637 < \sqrt{1/3}
\end{equation*}
implying that the initial condition is in the permitted set (\cref{sec:c_value}).

Since in practical applications randomness could seep into the system, we check robustness to measurement noise. To emulate measurement noise, white noise of relative magnitude, $ 10^{-3} $ is added to both the states $ (R, \Omega) $.

In this case too, convergence is observed to the desired equilibrium within the range of the measurement noise (\cref{fig:euc_rigid_noise}). We also notice that the state $ R $ is outside $ SO(3) $ initially, but soon converges to the manifold (modulo noise).

\section{Conclusions}
We initially introduced an existing linearization procedure for attitude control design in Euclidean space. Then, we proved that a single height function defined on the ambient Euclidean space, can be used to derive stabilizing nonlinear control for the attitude of a rigid body with a prescribed region of attraction. This is also illustrated through exemplary simulations. The algorithm is robust to measurement noise and numerical computation errors arising from digital implementation.

\bibliographystyle{ieeetr}
\bibliography{root.bib}

\begin{thebibliography}{10}

\bibitem{tsiotras1995new}
P.~Tsiotras, ``New control laws for the attitude stabilization of rigid
  bodies,'' in {\em Automatic Control in Aerospace}, pp.~321--326, Elsevier,
  1995.

\bibitem{shuster1993survey}
M.~D. Shuster, ``A survey of attitude representations,'' {\em Navigation},
  vol.~8, no.~9, pp.~439--517, 1993.

\bibitem{mortensen1968globally}
R.~E. Mortensen, ``A globally stable linear attitude regulator,'' {\em
  International Journal of Control}, vol.~8, no.~3, pp.~297--302, 1968.

\bibitem{junkins1991near}
J.~L. Junkins, Z.~Rahman, and H.~Bang, ``Near-minimum-time control of
  distributed parameter systems-analytical and experimental results,'' {\em
  Journal of Guidance, Control, and Dynamics}, vol.~14, no.~2, pp.~406--415,
  1991.

\bibitem{bhat2000topological}
S.~P. Bhat and D.~S. Bernstein, ``A topological obstruction to continuous
  global stabilization of rotational motion and the unwinding phenomenon,''
  {\em Systems \& Control Letters}, vol.~39, no.~1, pp.~63--70, 2000.

\bibitem{bloch2003nonholonomic}
A.~M. Bloch, ``Nonholonomic mechanics,'' in {\em Nonholonomic mechanics and
  control}, pp.~207--276, Springer, 2003.

\bibitem{bullo2004geometric}
F.~Bullo and A.~D. Lewis, {\em Geometric control of mechanical systems:
  modeling, analysis, and design for simple mechanical control systems},
  vol.~49.
\newblock Springer Science \& Business Media, 2004.

\bibitem{bayadi2014almost}
R.~Bayadi and R.~N. Banavar, ``Almost global attitude stabilization of a rigid
  body for both internal and external actuation schemes,'' {\em European
  Journal of Control}, vol.~20, no.~1, pp.~45--54, 2014.

\bibitem{crouch1984spacecraft}
P.~Crouch, ``Spacecraft attitude control and stabilization: Applications of
  geometric control theory to rigid body models,'' {\em IEEE Transactions on
  Automatic Control}, vol.~29, no.~4, pp.~321--331, 1984.

\bibitem{lee2011geometric}
T.~Lee, ``Geometric tracking control of the attitude dynamics of a rigid body
  on so(3),'' in {\em Proceedings of the 2011 American Control Conference},
  pp.~1200--1205, IEEE, 2011.

\bibitem{chang_feedback}
D.~E. Chang, F.~Jim{\'e}nez, and M.~Perlmutter, ``Feedback integrators,'' {\em
  Journal of Nonlinear Science}, vol.~26, no.~6, pp.~1693--1721, 2016.

\bibitem{arsie_patching}
A.~Arsie and C.~Ebenbauer, ``Locating omega-limit sets using height
  functions,'' {\em Journal of Differential Equations}, vol.~248, no.~10,
  pp.~2458--2469, 2010.

\bibitem{chang_controller}
D.~E. Chang, ``Controller design for systems on manifolds in {E}uclidean
  space,'' {\em arXiv preprint arXiv:1710.02780}, 2017.

\bibitem{chang2018controller}
D.~E. Chang, ``On controller design for systems on manifolds in {E}uclidean
  space,'' {\em International Journal of Robust and Nonlinear Control},
  vol.~28, no.~16, pp.~4981--4998, 2018.

\bibitem{murray2017mathematical}
R.~M. Murray, Z.~Li, and S.~Sastry, {\em A mathematical introduction to robotic
  manipulation}.
\newblock CRC press, 2017.

\end{thebibliography}
% \import{root.bbl}

\section{Appendix}
%\addtolength{\textheight}{3cm}   % This command serves to balance the column lengths

\subsection{Calculation of the parameter, $ c $}\label{sec:c_value}
We want to consider initial conditions for which the first state value does not lie on $ SO(3) $. To exactly verify the existence of such $ R(0) $ values, we need to evaluate the value of $ c $ in $ \tilde{V}^{-1}([0, c]) $ of \cref{thm:rigd_body_stab}.

We proceed by utilizing part of the proof of \cite[Lemma 2]{chang2018controller}. Define $ f : GL^{+}(3) \to \mathbb{R}_{\ge 0} $,
\[ f(R) = \frac{k_e}{4} ||R^T R - I||^2 \]
Take a small $ \delta > 0 $ such that every $ A \in \mathbb{R}^{3 \times 3} $ with $ ||A - I|| \le \delta $ is invertible. And let $ c = k_e \delta^2/ 4 $. Then, if $ R \in f^{-1}([0, c]), ||R^T R - I|| \le \delta $, meaning $ R^T R $ and $ R $ are invertible. Hence $ f^{-1}([0, c]) \subset GL^{+}(3) $. For a value of $ \chi $ close to $ \delta $,
\begin{gather*}
||A - I|| \le \delta < \chi
\Rightarrow \sum_{i=j} (A_{ij} -1)^2 + \sum_{i\neq j} A_{ij}^2 < \chi^2 \\
\Rightarrow (A_{ii} -1)^2 + \sum_{j \neq i, j=1}^{3} A_{ij}^2 < \chi^2
\end{gather*}
for any $ i = 1, 2, 3 $. So,
\begin{gather*}
0 > 2 (A_{ii}^2 - 2 A_{ii} + 1 - \chi^2 + \sum_{j \neq i} A_{ij}^2) \\
\Rightarrow A_{ii}^2 - 2 \sum_{j \neq i} A_{ij}^2 > 3 A_{ii}^2 - 4 A_{ii} + 2 (1 - \chi^2)
\end{gather*}
If $ 4^2 - 4 \times 3 \times 2 (1 - \chi^2) < 0 \Rightarrow \chi < \sqrt{1/3} $, RHS of the above equation is always positive. Hence,
\begin{gather*}
A_{ii}^2 - 2 \sum_{j \neq i} A_{ij}^2 > 0\\
\Rightarrow A_{ii}^2 > \sum_{j \neq i} A_{ij}^2 + 2 \prod_{j \neq i} |A_{ij}|\\
\Rightarrow |A_{ii}| > \sum_{j \neq i} |A_{ij}|
\end{gather*}
which means that the matrix $ A $ is strictly diagonally dominant. In summary, if $ \delta < \chi < \sqrt{1/3} $, $ A $ is invertible. Now, the next part of the proof of \cite[Lemma 2]{chang2018controller} is continued as is, to arrive at \cite[Theorem 2]{chang2018controller}.

Hence we obtain a sufficient condition that the permitted values of $ R(0) $ should satisfy,
\begin{gather*}
 \tilde{V}(R(0)) \le (c = k_e \delta^2 / 4) <  k_e/ 12 \\
 \Rightarrow ||R(0)^T R(0) - I|| < \sqrt{\frac{1}{3}}
\end{gather*}

\subsection{Calculation of derivative of the height function}\label{sec:deriv_height}
We need to evaluate the derivative of $ W(x) $ along the flow (Lie derivative) on the submanifold $ \mathcal{M} $. A few useful relations are,
\begin{itemize}
\item $ tr(A) = tr(A^T), tr(ABC) = tr(BCA) $; $ A, B, C $ are square matrices
\item $ tr(AB) = 0 $ if $ A $ is a symmetric matrix and $ B $ is a skew-symmetric matrix
\item $ \widehat{(v \times w)} = [\hat{v}, \hat{w}] = 2\ \text{skew}(\hat{v}\hat{w}) $, $ v, w \in \mathbb{R}^3 $
\item $ v^T w = \frac{1}{2} \langle \hat{v}, \hat{w} \rangle $, $ v, w \in \mathbb{R}^3 $; $ ||v||^2 = \frac{1}{2} ||\hat{v}||^2 $
\end{itemize}

Now, the system \eqref{eq:rigid_body_euc} restricted to the submanifold $ \mathcal{M} $ can also be written as,
\begin{equation} \label{eq:rigid_body_Z}
\begin{aligned}
\dot{Z}_s &= \frac{1}{2} (Z_s\hat{\Omega}-\hat{\Omega}Z_s) + Z_k\hat{\Omega} - \frac{1}{2} \widehat{Z_k^\vee \times \Omega} \\
\dot{Z}_k &= \frac{1}{2} (Z_s\hat{\Omega}+\hat{\Omega}Z_s) + \hat{\Omega} + \frac{1}{2} \widehat{Z_k^\vee \times \Omega} \\
\dot{\Omega} &= -k_p Z_k^\vee - k_d \Omega
\end{aligned}
\end{equation}
where $ Z = R_0^T (R - R_0) $. We have chosen the height function as,
\begin{gather*}
W(R, \Omega) = \frac{k_p}{4} (||Z_s||^2 + ||Z_k||^2) + \frac{1}{2}||\Omega||^2 + \epsilon \langle Z_k^\vee, \Omega \rangle
\end{gather*}
So its Lie derivative,
{\small \begin{multline*}
\dot{W}|_\mathcal{M}(R, \Omega) = \frac{k_p}{2}\left( \left\langle Z_s, \frac{1}{2} (Z_s\hat{\Omega}-\hat{\Omega}Z_s) + Z_k\hat{\Omega} - \frac{1}{2} \widehat{Z_k^\vee \times \Omega} \right\rangle \right. \\
\left. + \left\langle Z_k, \frac{1}{2} (Z_s\hat{\Omega}+\hat{\Omega}Z_s) + \hat{\Omega} + \frac{1}{2} \widehat{Z_k^\vee \times \Omega} \right\rangle \right) \\
+ \langle \Omega, -k_p Z_k^\vee - k_d \Omega \rangle + \epsilon \left\langle \frac{1}{2} (Z_s\hat{\Omega}+\hat{\Omega}Z_s)^\vee + \Omega + \frac{1}{2} Z_k^\vee \times \Omega, \Omega \right\rangle\\
+ \epsilon \langle Z_k^\vee, -k_p Z_k^\vee - k_d \Omega \rangle
\end{multline*}}
We know that,
\begin{gather*}
\langle Z_s, Z_s\hat{\Omega} \rangle = tr(Z_s^T Z_s \hat{\Omega}) = tr((Z_s^T Z_s) \hat{\Omega}) = 0 \\
\langle Z_s, \hat{\Omega}Z_s \rangle = tr(Z_s^T \hat{\Omega} Z_s) = tr(\hat{\Omega} (Z_s Z_s^T )) = 0 \\
\langle Z_s, \widehat{Z_k^\vee \times \Omega} \rangle = tr(Z_s^T (\widehat{Z_k^\vee \times \Omega})) = 0 \\
\langle Z_k, \widehat{Z_k^\vee \times \Omega} \rangle = 2 \langle Z_k^\vee, Z_k^\vee \times \Omega \rangle = 2 \langle \Omega, Z_k^\vee \times Z_k^\vee \rangle = 0 \\
\langle \Omega, Z_k^\vee \times \Omega \rangle = \langle \Omega \times \Omega, Z_k^\vee \rangle = 0 \\
\langle Z_k, \hat{\Omega} \rangle = 2 \langle Z_k^\vee, \Omega \rangle
\end{gather*}
{\small \begin{multline*}
\left\langle Z_k, \frac{1}{2} (Z_s\hat{\Omega}+\hat{\Omega}Z_s) \right\rangle = \frac{1}{2} tr(-Z_k Z_s \hat{\Omega} - Z_k \hat{\Omega} Z_s)\\
= \frac{1}{2} tr((-Z_k Z_s \hat{\Omega})^T - Z_k \hat{\Omega} Z_s) = \frac{1}{2} tr(-\hat{\Omega} Z_s Z_k - Z_k \hat{\Omega} Z_s)\\
= \frac{1}{2} tr(-Z_s Z_k \hat{\Omega} - Z_s Z_k \hat{\Omega}) = - tr(Z_s^T Z_k \hat{\Omega}) = - \langle Z_s, Z_k\hat{\Omega} \rangle
\end{multline*}}
{\small \begin{multline*}
\left\langle (Z_s\hat{\Omega}+\hat{\Omega}Z_s)^\vee, \Omega \right\rangle = \frac{1}{2} \left\langle \hat{\Omega}, (Z_s\hat{\Omega}+\hat{\Omega}Z_s) \right\rangle\\
= \frac{1}{2} tr(\hat{\Omega}^T Z_s \hat{\Omega} + \hat{\Omega}^T \hat{\Omega} Z_s) = tr(\hat{\Omega}^T Z_s \hat{\Omega})
\end{multline*}}
So we have the simplification,
\begin{multline*}
\dot{W}|_\mathcal{M}(R, \Omega) = -(k_d - \epsilon) ||\Omega||^2 - \epsilon k_d \langle Z_k^\vee, \Omega \rangle - \epsilon k_p ||Z_k^\vee||^2\\
+ \frac{\epsilon }{2} tr(\hat{\Omega}^T Z_s \hat{\Omega})
\end{multline*}
Now with $ \tilde{R} = R_0^T R $,
\begin{gather*}
x^T Z_s x = x^T ((R_0^TR - I)_s) x \\
= x^T (\tilde{R}_s - I) x = x^T (\frac{\tilde{R} + \tilde{R}^T}{2} - I) x \\
= 0.5 (x^T\tilde{R}x + x^T\tilde{R}^T x) - ||x||^2 \\
\le 0.5 (||x||^2 + ||x||^2) - ||x||^2 \le 0
\end{gather*}
as $ \tilde{R} $ is a rotation matrix and $ ||\tilde{R}x|| = ||x|| $. So $ Z_s $ is negative semi-definite implying,
\begin{gather*}
\quad tr(\hat{\Omega}^T Z_s \hat{\Omega}) = \sum_{i=1}^3 (\Omega \times e_i)^T Z_s (\Omega \times e_i) \le 0
\end{gather*}
with $ (e_i)_j = \delta_{i, j}, j= 1...3 $. Hence,
\begin{gather*}
\dot{W}|_\mathcal{M}(R, \Omega) \le -(k_d-\epsilon) ||\Omega||^2 - \epsilon k_d \langle Z_k^\vee, \Omega \rangle - \epsilon k_p ||Z_k^\vee||^2 \le 0
\end{gather*}
is negative definite if $\displaystyle 0 < \epsilon < \frac{4k_pk_d}{4k_p+k_d^2} $.

\end{document}